# Two-step frequency conversion for connecting distant quantum memories by transmission through an optical fiber

Shuhei Tamura,[1] Kohei Ikeda,[1] Kotaro Okamura,[2] Kazumichi Yoshii,[1] Feng-Lei Hong,[1] Tomoyuki Horikiri,[1,3,*] and Hideo Kosaka[1]

[1]*Department of Physics, Yokohama National University, 79-5 Tokiwdai Hodogaya, Yokohama, Kanagawa 240-501, Japan*

[2]*Research Institute of Engineering, Kanagawa University, 3-27-1 Rokkakubashi, Yokohama 221-8686, Japan*

[3]*JST, PRESTO, 4-1-8 Honcho, Kawaguchi, Saitama, 332-0012, Japan*

[*]E-mail: horikiri-tomoyuki-bh@ynu.ac.jp

Long-distance quantum communication requires entanglement between distant quantum memories. For the purpose, photon transmission is necessary to connect the distant memories. Here, for the first time, we develop a two-step frequency conversion process (from a visible wavelength to a telecommunication wavelength and back) involving the use of independent two-frequency conversion media where target quantum memories are nitrogen-vacancy centers in diamonds (with an emission/absorption wavelength of 637.2 nm), and experimentally characterize the performance of this process acting on light from an attenuated CW laser. A total conversion efficiency of approximately 7% is achieved. The noise generated in the frequency conversion processes is measured, and the signal-to-noise ratio is estimated for a single photon signal emitted by an NV center. The developed frequency conversion system has future applications via transmission through long optical fiber channel at a telecommunication wavelength for quantum repeater network.

# 1. Introduction

Long-distance quantum entanglement is required for quantum information processes such as quantum key distribution [1,2], quantum teleportation [3], and distributed quantum computation [4]. However, the efficient generation of entanglement between remotely separated quantum memories has not yet been achieved and remains quite challenging. Quantum memory that preserves the quantum state of an incoming photon as a static qubit in the memory material and re-emits a photon having the same quantum state as the incoming photon is an important component of quantum repeaters [5,6]. To date, various materials have been investigated as quantum memories, including atomic gases [7], trapped ions [8], and solid-state materials such as semiconductor quantum dots [9], nitrogen-vacancy (NV) centers in diamond [10-13], and rare-earth-ion-doped crystals [14]. The read and write wavelengths of most quantum memories are around the visible or near-infrared regime, at which the loss in an optical fiber is much greater than that at telecommunication wavelengths (approximately 0.2 dB/km). Therefore, quantum frequency conversion (QFC) between visible (or near-infrared) and telecommunication wavelengths is necessary to achieve long-distance quantum communication.

QFC experiments have been performed using a nonlinear optical process, with periodically poled $LiNbO_3$ (PPLN) waveguide devices mainly utilized as key components for performing sum-frequency generation (SFG) and difference-frequency generation (DFG). Several QFC experiments related to the long-distance quantum entanglement of quantum memories have been conducted, which have involved Rb atomic ensembles (DFG: 780 nm → 1.5 μm) [15-18], $Pr^{3+}$:$Y_2SiO_5$ (SFG: 1570 nm → 606 nm) [19], silicon-vacancy centers in diamond (DFG: 738 nm → 1557 nm) [20], NV centers in diamond (DFG: 637 nm → 1587 nm), [21,22] and quantum dots (DFG: 910 nm → 1560 nm [9,23] or 1610 nm [24], and 711 nm → 1313 nm [25]).

In the present study, we assume that the quantum memory material is NV centers in diamond. An NV center is a defect created when a carbon atom is replaced by a nitrogen atom, and a vacancy exists next to the nitrogen atom. The NV center has long been considered as promising for quantum memory owing to the current second-order spin coherence time of NV electronic and nuclear spins [12,26,27] and nearby isotopic impurity $^{13}$C nuclear spins which can be utilized as multiqubit registers [27-30]. Various quantum manipulation techniques have already been demonstrated by utilizing NV centers, including arbitrary quantum state preparation, one- and two-qubit operation, entanglement generation, and quantum teleportation [12-13,31-37]. Therefore, NV centers are expected to provide tools for quantum information processing and long-distance quantum communication. It has been demonstrated that NV electronic and nuclear spins can be entangled with emitted and absorbed photons with a wavelength of 637.2 nm [10-12]. Common long-distance quantum entanglement generation involves the use of entangled photon sources and Bell measurement, but when NV centers are employed, the scheme can be as shown in Fig. 1.

The wavelength of a photon emitted from one NV center (Node A) is converted from 637.2 nm to a telecommunication wavelength, and the photon is then sent to a distant NV center (Node B) through an optical fiber. After a second wavelength conversion from the telecommunication wavelength to 637.2 nm at Node B, the converted photon is absorbed at Node B. By performing a measurement for entanglement generation, long-distance entanglement between Nodes A and B can be achieved.

In Ref. 21 and 22, it was shown that the wavelength of a photon emitted by an NV center can be converted to a telecommunication wavelength with suppressing the noise, the former is about 40 kcts/s and the latter is about 300 cts/s; thus, NV centers are a promising candidate to realize long-distance communication by utilizing a pump laser with a wavelength of around 1 μm for QFC. To investigate the feasibility of long-distance quantum communication based on NV centers, we investigated two-step frequency conversion [38,39] by using two separate QFC devices connected by an optical fiber.

In this article, we present the results of our two-step frequency conversion experiment and an estimation of the noise caused by the conversion process. We discuss whether the obtained noise level is sufficiently lower than the signal level of a single photon emitted from an NV center and whether this conversion scheme can be applied to long-distance quantum communication based on NV centers.

## 2. Experimental methods

The experimental setup is shown in Fig. 2. The 637.2-nm signal light, imitating a photon emitted from an NV center, was emitted by a laser diode (LD, Thorlabs: HL63142DG) and mixed with a pump laser beam (1071 nm, fiber-amplified external cavity diode laser (ECDL), Sacher Lasertechnik) at a dichroic mirror DM1 (Thorlabs, DMLP950), following which it was coupled to a PPLN ridge waveguide (NTT Electronics, chip length $L$ = 48 mm, both the input and output facets are anti-reflection-coated for 637 nm, 1587 nm, and 1064 nm) by using an aspheric lens with a numerical aperture (NA) of 0.16 and focal length of 5.0 mm. The coupling efficiency was 0.31 for the 637.2-nm laser and 0.33 for the pump laser. The PPLN waveguide was set on a three-axis stage so that the coupling of the waveguide with the laser light could be adjusted. To satisfy the phase-matching conditions, we used a temperature controller (Thorlabs, TED-200C) and a Peltier module. To imitate the spectral linewidth of a photon emitted from an NV center (approximately 10 MHz), we introduced a Littrow ECDL system by applying a holographic grating after the collimating lens of the LD. The demonstrated linewidth was on the order of megahertz. This procedure was necessary since the original LD had a spectral width of 1 nm, while the phase-matching bandwidth of PPLN is approximately 40 GHz (approximately 0.05 nm), as shown in Fig. 3. The narrow 637.2-nm ECDL prevented the QFC efficiency from decreasing because of the spectral mismatch.

The pump laser was employed in the two conversion processes by using a beam splitter. To maximize the coupling efficiency of the PPLN waveguide, each laser was adjusted to the appropriate polarization by a half-wave plate (HWP). At the first-conversion PPLN waveguide (PPLN1, Waveguide width: 7.8 μm, quasi-phase-matching (QPM) pitch: 11.47 μm, T: 318 K), DFG (637.2 nm to the telecommunication wavelength 1573.2 nm) was performed. After applying the long-pass filter (LPF, Thorlabs, FEL1100, transmission 87.0%) set just after PPLN1, the conversion efficiency was measured using a power meter. The LPF removed the pump light, second-harmonic generation (SHG) of the pump light, and 637.2-nm light from the converted light. After filtering, the converted light was sent to the second PPLN waveguide (PPLN2, Waveguide width: 7.8 μm, QPM pitch: 11.45 μm, T: 336 K) through a 1 m single-mode fiber (Cutoff wavelength < 1500 nm, the coupling efficiency was about 0.5). The converted light and pump laser were coupled to PPLN2 by a dichroic mirror DM2 (Thorlabs, DMLP1180) and an aspheric lens with an NA of 0.16 and a focal length of 5.0 mm. The coupling efficiency was 0.35 for the converted light and 0.25 for the pump laser. Then, SFG (from the telecommunication wavelength 1573.2 nm to 637.2 nm) was performed. After PPLN2, a short-pass filter (SPF, Thorlabs, FES0800, transmission 86.2%) and dichroic mirror DM3 (Thorlabs, DMLP950) removed the pump light, SHG light, and telecommunication-wavelength light. Then, the second quantum conversion efficiency was measured. To measure the noise spectra, the two-step converted light was sent to a single-photon count module (SPCM, Perkin Elmer, SPCM-AQR14) in a black box after transmission through a spectrometer (Acton, Spectrapro 750, nominal wavelength resolution of approximately 30 GHz) and bandpass filter (BPF, Semrock, 634–641 nm transmission > 0.93).

## 3. Results and discussion

## 3.1 Frequency conversion and noise measurement

The QFC efficiency is given by the measured power and $\eta_{ext} = \frac{n_{out}}{n_{in}}$, where $n_{in}$ is the mean incident photon number of the PPLN waveguide and $n_{out}$ is the mean converted photon number. The QFC efficiencies of SFG and DFG satisfy the equation [15] $\eta_{ext} = \eta_{ext}^{max} \sin^2(L\sqrt{P_p \eta_{nor}})$, where $P_p$ is the pump power, $\eta_{nor}$ is the normalized conversion efficiency, and $L$ is the crystal length. The pump-power dependence of the QFC efficiency is shown in Fig. 4. The maximum QFC efficiencies of the first and second conversions (DFG and SFG) were 27.1% and 25.6%, respectively, which were achieved with a pump power of 500 mW at each waveguide. From these results, the maximum total two-step QFC efficiency was determined to be approximately 7%. Based on the measured coupling efficiency, the maximum internal QFC efficiency was estimated to be 87% for DFG and 73% for SFG. The internal QFC efficiencies are different because different waveguides were utilized for two-step conversion.

In the QFC process, considerable noise is induced by the pump laser [25,40-42]. Since the single-photon count rate from an NV center is on the order of 1 Mcts/s [44], though it is performed by using a NV center in nanodiamonds where the single photon nature of the signal from the NV center was demonstrated by the second-order autocorrelation function $g^{(2)}(0)$ of 0.16, reducing the noise count rate is significant for enabling long-distance quantum communication. To evaluate the noise generated in this conversion scheme, we measured the noise level near 637.2 nm.

The noise spectrum around 637.2 nm obtained from the SPCM count rate is presented in Fig. 5. The spectrum of the 637.2-nm signal light appears in red, while that obtained with only the pump laser by blocking the signal light in front of the PPLN waveguide is shown in blue. The black plot represents the noise of the SPCM (dark count and/or stray light). The 637.2-nm laser intensity is approximately 1 nW. It is much stronger than the actual NV signal count rates around 1 MHz and is provided for reference. In this experiment, the output slit width of the spectrometer corresponded to a spectral width of 0.05 nm (40 GHz), and the bandwidth of the BPF inserted after the spectrometer was 7 nm. The noise spectrum agrees with the signal light spectrum. This is due to the resolution of the spectrometer.

It is known that the noise generated during frequency conversion mainly originates from the pump laser, and the main factors contributing to noise are Raman scattering and spontaneous parametric down-conversion (SPDC) [25,40-42].

Figure 6 illustrates how the noise originating from the pump laser is mixed with the signal in the two-step QFC process. In PPLN1, the pump laser (blue) generates considerable noise (green) around the telecommunication wavelength band because of SPDC and Raman scattering. SFG (red) of the telecommunication-wavelength noise light and pump laser can also occur. SFG light occurring in the first step can be blocked by the LPF, but the telecommunication-wavelength noise light is transmitted through the filter because its wavelength is the same as that of the first-converted telecommunication-wavelength light. Therefore, the noise light is sent to PPLN2 through the optical fiber. Most of the telecommunication-wavelength noise light generated by the pump laser is blocked by the SPF, but the residual noise light is converted to 637.2 nm via SFG and transmitted through the filter because its wavelength is the same as that of the light after the second conversion, 637.2 nm.

Figure 7 depicts the noise spectra induced by the pump laser. When the pump laser couples only to PPLN1 (yellow), no noise exists around 637.2 nm. The telecommunication-wavelength noise light generated by the pump laser in PPLN1 is partly converted to 637.2 nm (SFG of the noise light and pump light) and removed by LPF. The remaining telecommunication-wavelength noise is transmitted through the optical fiber and then enters PPLN2. However, the noise light is not converted to 637.2 nm, because the pump laser is off in the second conversion step.

When the pump laser couples only to PPLN2 (green), the SFG of the noise light and pump light occurs in the same spectrum as the PPLN phase-matching bandwidth. The blue points in Fig. 7 correspond to the case in which the pump laser couples with both PPLN1 and PPLN2. The noise is more significant in this case because the telecommunication-wavelength noise light from PPLN1 is mixed with that from PPLN2, and the SFG of the noise light and pump light occurs in PPLN2. The transmission rate from PPLN1 to PPLN2 is approximately 0.4, and the noise count in the case of coupling with both PPLN1 and PPLN2 (blue) is approximately 40% greater than that in the case of coupling with only PPLN2 (green).

## 3.2 SNR estimation

Here, we address the possibility of sending a single photon between distant NV centers by two-step conversion through an optical fiber as well as the absorption of the signal light and, consequently, long-distance entanglement generation between the two NV centers without interruption by noise.

In the present study, the single-photon detector count rate of the noise generated in the two-step frequency conversion process was on the order of 100 kcts/s. The spectral width of the noise was 40 GHz around a wavelength of 637.2 nm, as discussed above, owing to the phase-matching conditions of the PPLN waveguides. The noise spectrum due to SPDC and Raman scattering was broad across the telecommunication wavelength band, but the spectral width of the 637.2-nm noise light after PPLN2 was reduced to the PPLN waveguide phase-matching bandwidth in the SFG process.

As the linewidth of a photon from an NV center is approximately 10 MHz, the noise level can be decreased by three to four orders of magnitude by spectral filtering. The noise can be reduced to the order of 10 cts/s in the optimal case. Since we used a diffraction grating spectrometer in this study, the signal and noise count rate was reduced by one order of magnitude because of the low first-order diffraction efficiency. Therefore, in actual quantum communication between two distant NV centers, a spectrometer based on gratings should not be used. The combination of filters and Fabry-Perot cavities can reduce the noise to the order of 100 cts/s.

The photon count rate from an NV center is the order of 1 Mcts/s [44]; however, the rate will decrease by one order of magnitude after the present conversion process, which has an overall efficiency of approximately 7%. In this case, an ideal noise-filtering system will produce a signal-to-noise ratio (SNR) of 100 [kcts/s] / 100 [cts/s] ≈ 1000. To achieve a higher SNR, a higher QFC efficiency would be necessary. Furthermore, the distance between quantum repeater nodes is on the order of 10 km [45], and the photon transmission rate is reduced to 1/10 of its original value by a 50-km optical fiber with a loss of 0.2 dB/km. Therefore, if the node distance is set to 50 km, the SNR becomes approximately 100. However, the node distance in a recent quantum repeater scheme was on the order of kilometers [46] such that the optical fiber loss is less than 2 dB. Then, an SNR of 1000 could be achieved.

**3.3 Future applications and system stability**

To reach almost unit conversion efficiency, it is necessary to improve the efficiency of coupling to PPLN waveguides, which was the main limiting factor affecting the present study. This efficiency can be substantially improved by using a waveguide for which the input is fiber pigtailed [47,48], thereby improving the total conversion efficiency.

In the present two-step conversion study, we used a single pump laser for both DFG and SFG. However, to generate entanglement between distant quantum memories, it is practical to use different pump lasers since pump-laser power degradation can be avoided by setting each pump laser close to each quantum memory. Degradation would occur if only a single pump laser were used in such a situation because of the difficulty in maintaining the pump-laser power above 100 mW (thus attaining the maximum conversion efficiency) after several kilometers of optical-fiber transmission due to the optical fiber loss of approximately 1 dB/km at wavelengths of around 1 µm.

Another problem is that the photon emission and absorption wavelengths of NV centers experience detuning (on the order of gigahertz). We consider the case in which a pump laser beam is split into two, with one part used for the first NV center and the other used for the second NV center. The beam for the first NV center is used for DFG, while that for the second NV center is sent to a distant quantum memory for QFC (SFG). However, the pump laser cannot be used for both memories, because the converted 637.2-nm light from the telecommunication-wavelength signal photon and pump laser is detuned from the second NV absorption wavelength. To eliminate the detuning gap, one solution would be to use pump lasers with two different wavelengths.

One stability issue which may lead to a low conversion efficiency and a low coupling efficiency between two NV centers is frequency shifts of the lasers. The degree of frequency shift in the present work around 1 GHz was within the PPLN phase matching bandwidth (~ 50 GHz) and did not affect the efficiency. However, to connect distant NV centers in this scheme, it is necessary to stabilize the frequency of the pump lasers. A pump laser with a wavelength of around 1 µm can be locked to one absorption line of molecular iodine gas through the SHG of the laser and saturated absorption spectroscopy. A frequency stability much better than the megahertz order can be achieved by employing this scheme since the Doppler-free spectroscopy attainable using this method enables the resolution of hyperfine levels having narrow linewidths (on the order of megahertz), which are usually hidden in the Doppler-broadened linewidth on the order of gigahertz [49]. However, two pump lasers cannot be locked to the same absorption line because of detuning. There are two solutions to this problem. The first is that, if the NV detuning is sufficiently small (≲ 100 MHz), the frequency of the signal can be modulated after the two-step conversion process by an acousto-optic modulator (AOM). The second is that, if the detuning is larger than the limit of the frequency shift of the AOM, different iodine absorption lines can be used to lock the two pump lasers. There are 15 or 21 hyperfine levels in a single Doppler-broadened line (on the order of gigahertz). One can be employed for locking to achieve frequency stabilization below the order of kilohertz [50]. Selecting a different hyperfine structure line or changing the absorption line can provide a means of overcoming the issue of detuning greater than the gigahertz order. Finally, an AOM can be utilized to set the frequency to the second NV absorption wavelength because the iodine lines are discrete.

Another issue we need to think is polarization stability. In this experiment, the conversion efficiency stayed at its maximum due to a stable pump-laser power. However, polarization of the pump laser which is an output from a fiber amplifier (KEOPSYS, KPS-STD-BT-YFA-50-SLM-COL) is not stable for long-term, for example, a few hours. Therefore, we sometimes needed to optimize the polarization for obtaining the maximum conversion efficiency. In the laser utilized in Ref. [49], the polarization is also stable and the readjustment of polarization necessary in the present work can be avoided.

When connecting distant NV centers by using long optical fiber, it is also necessary to maintain the polarization qubits, because the emitted photon and spin state are entangled. However, decoherence of polarization qubits occurs easily owing to fiber bending and stress. Therefore, we need conversion of qubits from polarization to time-bin which is suitable for long distance communication [51,52].

## 4. Conclusions

In the present study, we demonstrated two-step frequency conversion by utilizing two independent frequency converters. Although we used a CW laser to imitate the NV center, the SNR estimation obtained from the total conversion efficiency shows that it is feasible to realize long-distance quantum communication over 100 km. And better QFC efficiencies utilizing a fiber-pigtail structure and the obtained pump-laser-induced noise level point to the possibility of connecting remote quantum memories through the one-way transmission of a single photon as well as the generation of distant entanglement between solid-state qubits. We believe that this study opens the possibilities for the realization of the NV centers based quantum repeater.

## Acknowledgments

We thank R. Ikuta, Y. Yamamoto, S. Utsunomiya, T. Kobayashi, S. Inoue, N. Namekata, M. Fraser, I. Iwakura, M. Zheng, X. Xie and Q. Zhang for their helpful comments and support.

This research was supported by Toray Science Foundation; the Asahi Glass Foundation; KDDI Foundation; Murata Science Foundation; JKA; REFEC; SECOM Foundation, JSPS Grant-in-Aid for Scientific Research (24244044, 16H06326, 16H01052), JST CREST, JST START ST292008BN, and JST PRESTO JPMJPR1769.

# Figure Captions

**Fig. 1.** (Color online) Long-distance entanglement creation between two diamond quantum memories.

**Fig. 2.** (Color online) Schematic of experimental setup. BS: Beam splitter, DM: Dichroic mirror, HWP: Half-wave plate, LPF: Long-pass filter, SPF: Short-pass filter, L1–L6: Lenses, BPF: Bandpass filter, SMF: Single-mode fiber, SPCM: Single-photon counting module.

**Fig. 3.** (Color online) Spectral acceptance bandwidth of PPLN waveguide. The vertical axis represents the DFG power when the spectrometer wavelength is scanned (horizontal axis), while the power of the 637.2-nm laser is constant.

**Fig. 4.** (Color online) Pump-laser-power dependence of QFC efficiency. (a) DFG (green), SFG (red). (b) Two-step conversion efficiency. The solid curves are fittings of the experimental data with the equation $\eta_{ext} = \eta_{ext}^{max} \sin^2(L\sqrt{P_p \eta_{nor}})$ for DFG (green) and SFG (red), while $\eta_{ext} = \eta_{ext}^{DFG} \times \eta_{ext}^{SFG}$ for Two-step (yellow).

**Fig. 5.** (Color online) Measured (a) signal (red), noise (blue), and dark count (black). (b) Reference noise spectrum around 637.2 nm.

**Fig. 6.** (Color online) Noise generation process. Pump laser (blue), noise light around the telecommunication wavelength band (green), and SFG of the noise and pump light (red).

**Fig. 7.** (Color online) Noise spectra induced by pump laser. Circles (blue): pump laser couples to both PPLN1 and PPLN2; diamonds (green): pump laser couples only to PPLN2; squares (yellow): pump laser couples only to PPLN1.

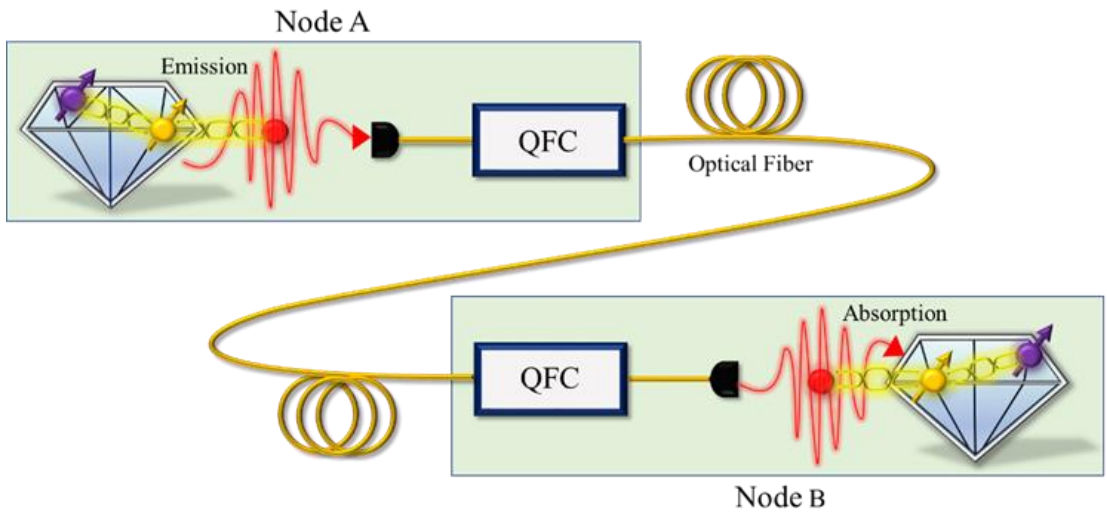

Fig.1.   (Color Online)

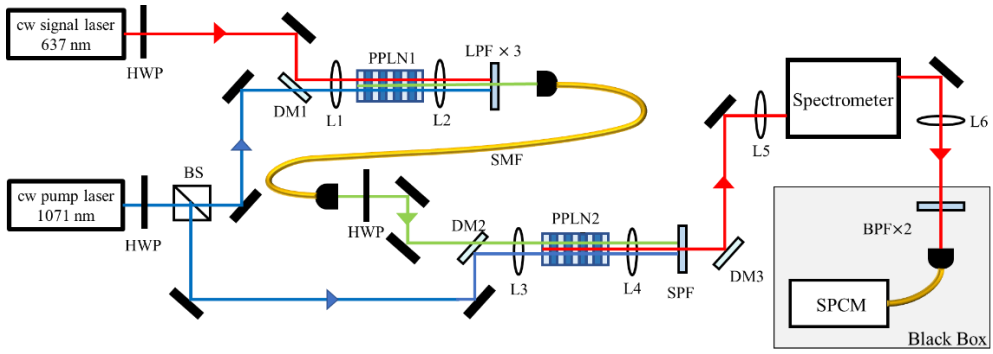

Fig.2. (Color Online)

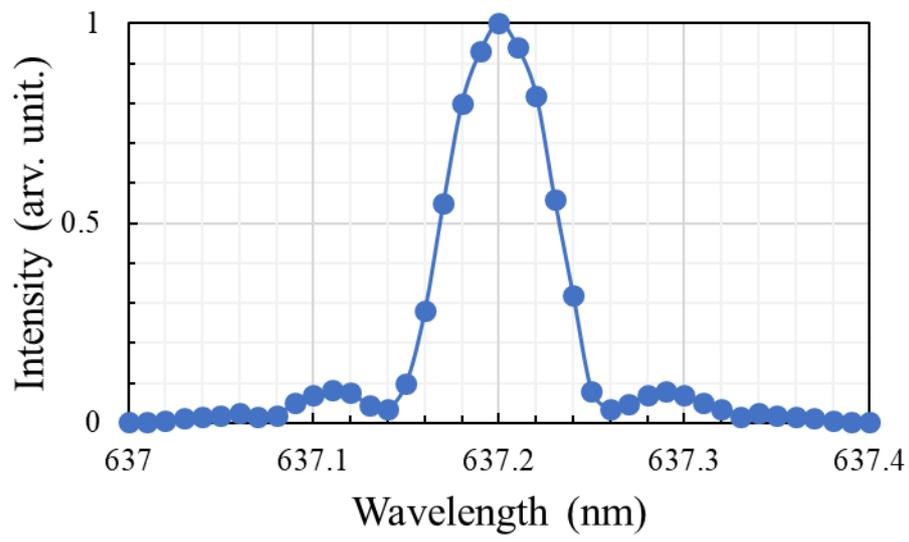

Fig.3. (Color Online)

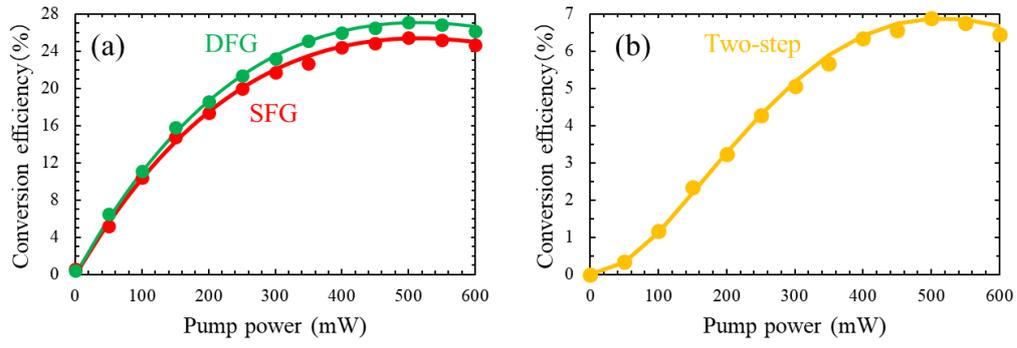

Fig.4.  (Color Online)

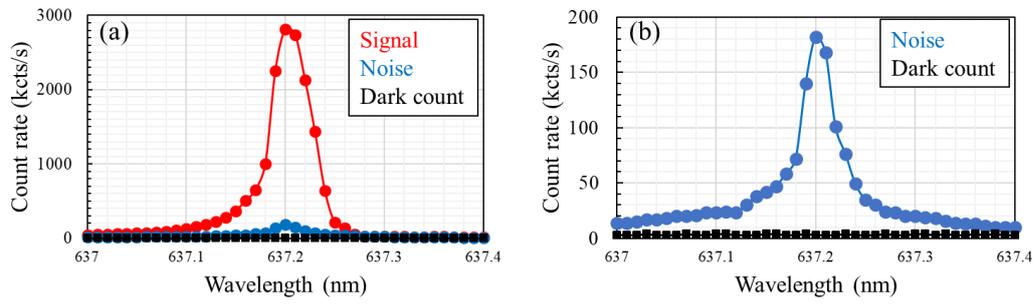

Fig.5. (Color Online)

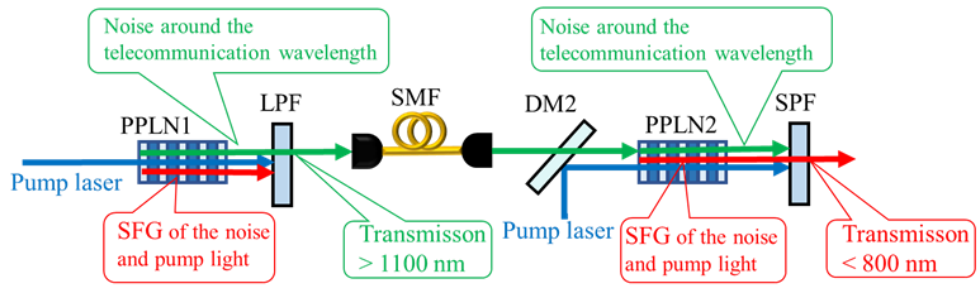

Fig.6. (Color Online)

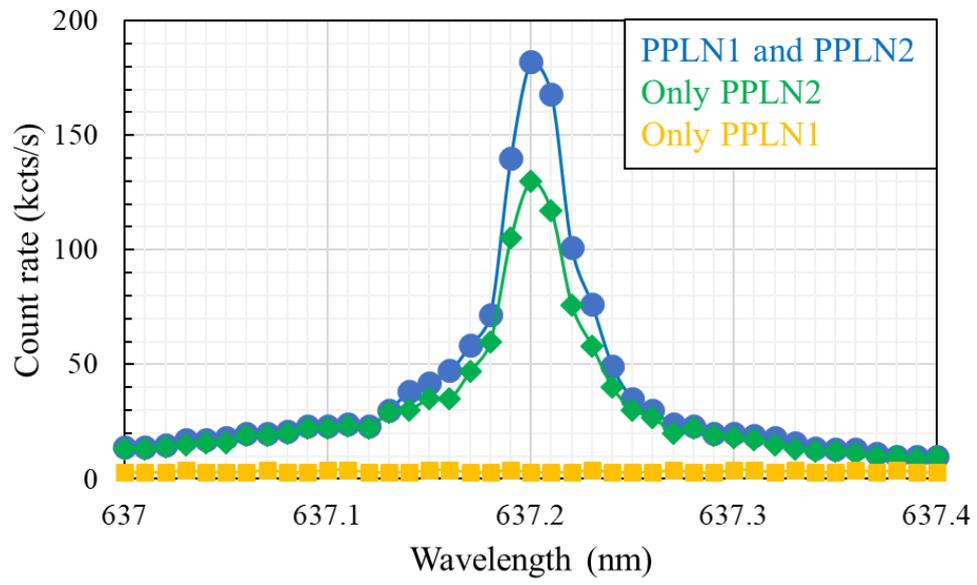

Fig.7.　(Color Online)